\documentstyle[aps,prl,multicol,epsf]{revtex}
\renewcommand{\narrowtext}{\begin{multicols}{2} \global\columnwidth20.5pc}
\renewcommand{\widetext}{\end{multicols} \global\columnwidth42.5pc}
\renewcommand{\v}[1]{{\bf #1}}

\newcommand{\ba}{\begin{eqnarray}}
\newcommand{\ea}{\end{eqnarray}}
\newcommand{\be}{\begin{equation}}
\newcommand{\ee}{\end{equation}}
\newcommand{\nn}{\nonumber\\}
\newcommand{\Eq}[1]{Eq.~(\ref{#1})}

\begin{document}
\draft

\title{Pairing in the spin sector}
\author{Qiang-Hua Wang$^{a,b}$,Jung Hoon Han$^{a}$,
and Dung-Hai Lee$^{a}$}
\address{${(a)}$Department of Physics,University of California
at Berkeley, Berkeley, CA 94720, USA \\}
\address{${(b)}$ Physics Department and National Laboratory
of Solid State Microstructures,\\ Institute for Solid State Physics,
Nanjing University, Nanjing 210093, China\\}

\maketitle \draft
\begin{abstract}
The nanometer-scale gap inhomogeneity revealed by recent STM
images of BSCCO surface suggests that the ``gap coherence length''
is of that order. Thus a robust pairing gap can develop despite
the poorly screened Coulomb interaction. This can be taken as an
evidence that the pairing in high $T_c$ materials occurs primarily
in the neutral (spin) sector. We provide theoretical support for
this point of view.
\end{abstract}

\narrowtext

It was first emphasized to us by Pan and Davis\cite{private} that
the energy gap extracted from STM measurements of BSCCO surface is
inhomogeneous at nanometer scale. Moreover this is true for
systems ranging from under- to slightly
over-doping.\cite{kapitulnik,pan,lang} Pan conjectured that such
inhomogeneity is due to the variation of the carrier density in
the copper-oxygen plane caused by randomly positioned dopant
oxygen.\cite{pan}

The above findings suggest that the ``gap coherence length'' of high-$T_c$
materials is at most a few nanometer. Since at such short length scale the
Coulomb interaction is poorly screened, it must be true that the pairing
takes place primarily in the neutral (spin) sector\cite{anderson} and
hence hardly affects the charge density correlation.

Ever since the BCS theory, superconductivity has been attributed
to the pairing of electrons. In the case of high-$T_c$
superconductors, it is sometimes  stated that superconductivity
requires the binding of doped holes. Conceptually it is important
to distinguish binding from pairing. The former is a feature in
density-density correlation while the latter is manifested in
two-particle off-diagonal correlation.

The distinction between pairing and binding is particularly pertinent in
the cuprates because of the short gap coherence length. (By gap coherence
length we mean the minimum length required for the pairing gap to be
exhibited.) Based on the STM results we argue that such length is of order
of nanometers. This in turn suggests that the pairing responsible for
high-$T_c$ superconductivity takes place primarily in the spin degrees of
freedom.

We support our point of view by first demonstrating that despite
strong Coulomb interaction in the following Hamiltonian
\begin{eqnarray}
H&=&-t\sum_{\langle ij\rangle} (c_{j\alpha}^\dag c_{i\alpha}
\!+h.c.) +J\sum_{\langle ij \rangle} (\v S_i\cdot\v
S_j \! -\frac{1}{4}\!n_{i}n_{j})\nonumber \\
 &+& {V_c}\sum_{i>j}
 {1\over {r_{ij}}}(n_{i}-\bar{n})(n_{j}-\bar{n}),
\label{h}
\end{eqnarray}
it is energetically favorable for the following variational
ansatz\cite{anderson} to develop  d-wave pairing: \be
|\Psi(D,\Delta)\rangle =P_G |\Psi_{mf}(D,\Delta)\rangle. \label{ans} \ee
Here $P_G$ is the projection operator that removes double occupancy and
$|\Psi_{mf}(D,\Delta)\rangle$ is the ground state of the following
mean-field Hamiltonian \ba H_{mf}(D,\Delta)&&=\sum_k[X_k
c_{k\sigma}^{\dagger}c_{k\sigma}+iD_k
c_{k+Q,\sigma}^{\dagger}c_{k\sigma}\nn&& +(\Delta_k
c_{k\uparrow}^{\dagger}c_{-k\downarrow}^{\dagger}+{\rm h.c.})],\ea where
$X_k=-2(\cos k_x+\cos k_y)$, $D_k=2D(\cos k_x -\cos k_y)$, and
$\Delta_k=2\Delta(\cos k_x -\cos k_y)$. In the rest of the paper we use
$t/J=3$.

By a straightforward minimization we conclude that for a wide doping range
it is energetically favorable to develop a non-zero $\Delta$ (not $D$) for
$V_c$ as big as $9J$. This result implies that a non-zero $\Delta$ hardly
perturb the density-density correlation, hence cannot cause hole binding.
We emphasize that the purpose of the present discussion is not to argue
that \Eq{ans} is the ground state of \Eq{h}, rather it is to show that the
pairing exhibited by \Eq{ans} is not hole binding.

To appreciate the effect of the projection operator in \Eq{ans} we have
also investigated the stability of $|\Psi_{mf}(0,\Delta)\rangle$ without
the Gutzwiller projection. In a nearest-neighbour repulsion model the
analytic result suggests that once $V_{nn}$, the nearest-neighbour
interaction strength, is larger than $\approx 0.5J$, pairing is absent!

Now we provide the details. We minimize
$E(D,\Delta)=\langle\Psi|H|\Psi\rangle/\langle\Psi|\Psi\rangle$ by
varying $D$ and $\Delta$. The results presented below are obtained
for a system with the average number of holes per site ($x$) equal
to $12\%$. The evaluation of $E(D,\Delta)$ is performed by Monte
Carlo when necessary.

In Fig.1(a) we present the projected results for $\Delta E\equiv
E(0,\Delta)-E(0,0)$ versus $\Delta$ in a $10\times 10$ lattice with 12
holes. The blue open circles are for the pure t-J model, the black crosses
are for t-J model with a nearest neighbor repulsion $V_{nn}=3J$, and the
red open squares are for t-J + Coulomb model (\Eq{h}) with $V_c=3J$. For
each of the three cases a nonzero  $\Delta$ develops.

In  Fig.1(b) we present the corresponding results for the
nearest-neighbor repulsion model when the projection operator is
removed from the variational ansatz. As one can see even for
$V_{nn}=0$ the optimal value of $\Delta$ is suppressed. Moreover
for $V_{nn}\geq 0.5J$ the optimal $\Delta$ is zero.

Since the presence/absence of spontaneous staggered current order is a
timely issue,\cite{ddw} we have also studied the optimum form of \Eq{ans}
allowing both $D$ and $\Delta$. In Fig.1(c) we plot $\Delta
E=E(D,\Delta)-E(0,0)$ for the t-J+Coulomb model ($V_c=3J$) at $x=12\%$. It
is clear from this plot that the minimum corresponds to a non-zero
$\Delta$ but $D=0$. Consequently we conclude that for the doping relevant
to the present paper d-wave pairing is the only order that occurs for the
model described by \Eq{h}. \widetext
\begin{figure}
\centering \epsfxsize=15cm\epsfysize=5cm\epsfbox{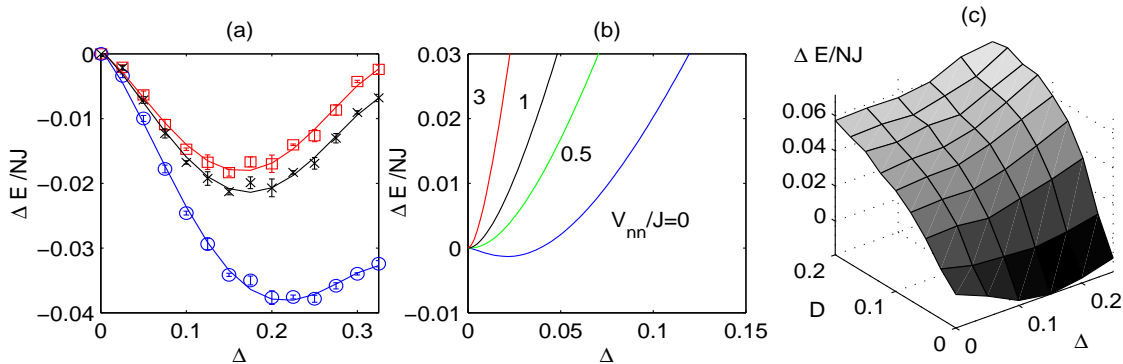}
\caption{Results for $x=12$\%. (a) $\Delta E$ as a function of $\Delta$ at
$D=0$ from Gutzwiller projection in a $10\times 10$ lattice. Open circles:
pure t-J model; crosses: t-J model with nearest neighbor repulsion
$V_{nn}=3J$; squares: t-J model with Coulomb interaction $V_c=3J$. (b)
Without the Gutzwiller projection, analytic results for nearest-neighbor
model. c) $\Delta E(D,\Delta)$ as a function of $D$ and $\Delta$ for the
projected wavefunction at $V_c=3J$.}
\end{figure}
\narrowtext

The next  question is, when subjected to disorder potential,
whether the wavefunction in \Eq{ans} can adjust its pairing
parameter $\Delta$ ``adiabatically'' to the local density to
account for the observed gap inhomogeneity. Unfortunately with the
disorder potential breaking the translation symmetry the vast
variational freedom renders variational Monte-Carlo impossible.
What we shall do in the rest of the paper is slave-boson
mean-field theory which takes the projection operator in \Eq{ans}
into account in an average fashion. The hope is that since such
mean-field theory qualitatively captures the physics of \Eq{ans}
in the absence of disorder it will produce meaningful result in
its presence.

The starting point of our mean-field theory is the following
Hamiltonian\cite{kotliarliu}
\begin{eqnarray}
H&=&-t\sum_{\langle ij\rangle} (b_jb^{\dag}_i f_{j\alpha}^\dag
f_{i\alpha} \!+h.c.) + J\sum_{\langle ij \rangle}
(\v S_i\cdot\v S_j \! -\frac{1}{4}\!n_{i}n_{j})\nonumber \\
 &+& {V_c}\sum_{i>j}
 {1\over {r_{ij}}}(n_{i}-\bar{n})(n_{j}-\bar{n})+\sum_i U_i n_i.
 \label{h1}
\end{eqnarray}
In the above $b$ and $f$ are the holon and spinon operators
obeying $f^\dag_{i\alpha}f_{i\alpha}+b^\dag_i b_i=1$,
$n_i=1-b^\dag_{i}b_{i}$, and $\v
S_i=(1/2)f^\dag_{i\alpha}\vec{\sigma}_{\alpha\beta}f_{i\beta}$.
The disorder potential $U_i$ is given by \be U_i =
\sum_{imp=1}^{N_{imp}} {V_d \over\sqrt{|\v r_i \!-\!\v r_{imp} |^2
\!+\! d^2 }}. \ee In the above $\v r_{imp}$ is the location of the
oxygen dopant and $d$, the setback distance, is the vertical
separation of the oxygen dopant plane from the copper-oxygen
plane.\cite{note} From simple valence counting we expect $N_{imp}$
to be half the number of holes. The results reported in the rest
of the paper are obtained on $24\times 24$ lattice for doping
$x=12\%$ using $V_d/2=V_c=t=3J$, and $d=2\times$ the CuO$_2$
lattice constant. We have checked that the results change smoothly
with $d$ and $x$.

The mean-field trial wavefunction $|\psi \rangle$ is given by $|\psi
\rangle=|\psi_b \rangle \otimes |\psi_f \rangle$ where the bosonic
$(|\psi_b \rangle)$ and fermionic ($|\psi_f \rangle)$ states are \ba
&&|\psi_b \rangle =[\sum_j\chi_j b^\dag_j]^{N_b}|0_b \rangle\nn &&|\psi_f
\rangle=\prod_{n} [\sum_j(u^*_{nj}f^\dag_{j\uparrow}+
v^*_{nj}f_{j\downarrow})]
 |0_f \rangle. \label{ans1} \ea  The
mean-field single-particle wavefunctions $u_{nj}, v_{nj}$ and $\chi_j$ are
varied to minimize \ba
 \langle\Psi|H|\Psi \rangle&-&\sum_i\lambda_i \langle\Psi|b^{\dag}_ib_i+
f^{\dag}_{i\alpha}f_{i\alpha}-1|\Psi \rangle\nn &-&\mu\sum_i\langle \Psi|
n_i -\bar{n} |\Psi \rangle . \label{qu} \ea Lagrange multipliers
$\lambda_i$, and $\mu$ are introduced to guarantee that the {\it average
occupation} obeys the constraints locally as well as globally. {\it
Assuming} that the inhomogeneity in STM images are indeed induced by the
dopant disorder, then, this model should capture the essential features of
what is seen on BSCCO surface.

The biggest difference between screening in an ordinary metal and
in the cuprates is that the latter is close to the Mott-insulating
limit. Due to the no-double-occupancy constraint the ability for
charge to redistribute is severely hindered. For example while the
constraint has no effect on the electron depletion, it does forbid
local electron accumulation beyond one electron per site. As a
consequence, in the distribution of holes there can be local peaks
of  hole density while the opposite, i.e. sharp local depletion of
holes, tends not to occur. This is indeed what comes out of our
calculation.

In Fig. 2 we show the bare impurity potential $U_i$ (Fig. 2(a))
and the corresponding hole distribution (Fig. 2(b)). By comparing
the two figures, one sees that the hole distribution does
correlate with the bare potential.

\begin{figure}
\hskip -0.7cm \epsfxsize=9cm\epsfysize=3.5cm \epsfbox{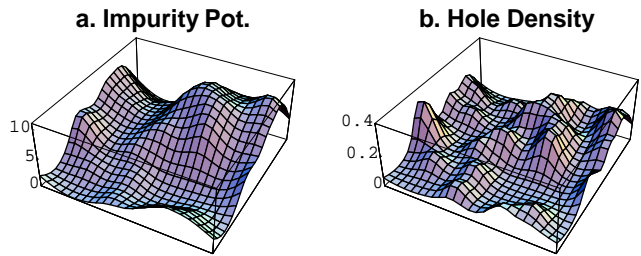}
\caption{A plot of the (a) bare impurity potential $U_i$  and (b)
the hole distribution (b) in a $24\times 24$ lattice with $x=12\%$
and $d=2$.}
\end{figure}

We now discuss the mean-field prediction of the tunneling
spectroscopy. The local differential conductance measured at a
bias $V$ by STM is proportional the electron local density of
states. In the slave-boson theory if one writes $b_i=|b_i|\phi_i$
($\phi$ is a U(1) phase factor) and ignore the fluctuation in
$|b_i|$, the electron spectral function is given by \be \rho_{e,i}
(V)=|b_i |^2 \rho_{qp,i}(V),\label{tdos0}\ee  where
$\rho_{qp,i}(V)$ is the local spectral function of the {\it
quasiparticles} and $|b_i|^2$ is the local hole density. (The
quasiparticle creation operator is given by $\phi_i
f^\dag_{i\alpha}$.) In the mean-field theory where the holon phase
fluctuation is ignored one obtains \be \rho_{qp,i}(V)=\sum_n
|u_{ni}| ^2 \delta (V\!\!-\!\!E_n ) \!+\! \sum_n |v_{ni}|^2 \delta
(V\!\!+\!\!E_n ). \label{tdos} \ee In the above  $(u_{ni},v_{ni})$
is the Bogoliubov-deGennes eigenfunctions of the  spinon
self-consistent mean-field Hamiltonian.

From \Eq{tdos} it is clear that when integrated over energy $V$,
the quasiparticle spectral function obeys a sum rule,\be
\int_{-\infty}^{\infty}\rho_{qp,i}(V)dV=2, \ee  hence is
independent of the site index $i$. This is not true for the
electron spectral function. Due to the presence of the $|b_i|^2$
factor the total integrated value of the electron spectral
function depends on the local hole density and hence varies from
site to site! Such lack of spectral conservation is a generic
property of a Mott insulator\cite{hanlee}. Equation (\ref{tdos0})
suggests that by properly dividing out the integrated electron
density of states (and hence $|b_i|^2$) one can get the
quasiparticle density of states.

It turns out that in the actual experiment this is customarily
done. In an STM experiment it is common to have the feedback loop
set up so that the total tunneling current (i.e. the integral of
the differential conductance up to a voltage V$_{max}$) is held at
a fixed value\cite{pan,lang} This way of calibration  precisely
divides out $|b_i|^2$ in \Eq{tdos0}. Thus the tunneling spectra
presented in Ref.\cite{lang} is, in our language, the
quasiparticle local density of states. One can also undo the
calibration to restore the $|b_i|^2$ and hence obtain the electron
local density of states\cite{pan}. It turns out that these two
density of states have interesting observable differences.

In Figure 3, we plot the quasiparticle ($\rho_{qp,i}(V)$) and the
electron ($\rho_{e,i}(V)$) local density of states in the bias
range of $-0.6J\le V\le +0.6J$ for four different locations in
Fig. 2. Among the four curves the local hole density ($|b_i|^2$)
varies from $0.055$ to $0.174$. As one can see, the peak-to-peak
distance, i.e. the local gap, varies considerably among the
curves. \cite{pan,lang,kapitulnik}.

Let's now focus on the behavior of $\rho_{qp,i}(V)$, and
$\rho_{e,i}(V)$ at small $V$. While different $\rho_{qp,i}(V)$
curves tend to merge at small voltage, the $\rho_{e,i}(V)$ curves
do not. This difference is precisely caused by the fact that each
curve has a different $|b_i|^2$. It is amazing that this
difference is seen in the experimental curves by changing the
calibration!\cite{pan,lang}

\begin{figure}
\hskip -0.2cm\centering \epsfxsize=8.5cm \epsfysize=3cm
\epsfbox{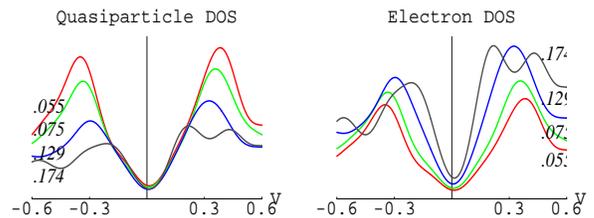} \caption{Tunneling conductance of the fermions
($\rho_{s,i}(V)$) and of electrons ($\rho_{e,i}(V)$) at several positions
in the $24\times24$ lattice (same as in Fig. 1). The local hole densities
corresponding to each curve is indicated in the figure. The bias voltage
$V$ is in units of $J$.}
\end{figure}

What is the reason for the universality of the quasiparticle
density of states at low energy? The answer is the robustness of
the low-energy quasiparticle wavefunctions due to the vanishing
density of states. In Fig. 4 we show $|v_{ni}|^2$ for two of the
low-lying energy eigenstates (associated with two different nodes)
from two different disorder realization, (a)-(b) and (c)-(d)
respectively. These eigenstates show a simple, geometric pattern,
insensitive to the underlying disorder. The orientation of the
wavefunction is also consistent with the direction of the
wavevector of the nodal quasiparticles. Such regularity persists
up to about $0.15J$, the same energy scale below which the
quasiparticle density of states appears universal.

\begin{figure}
\hskip -0.3cm \centering \epsfxsize=8.5cm \epsfysize=7cm
\epsfbox{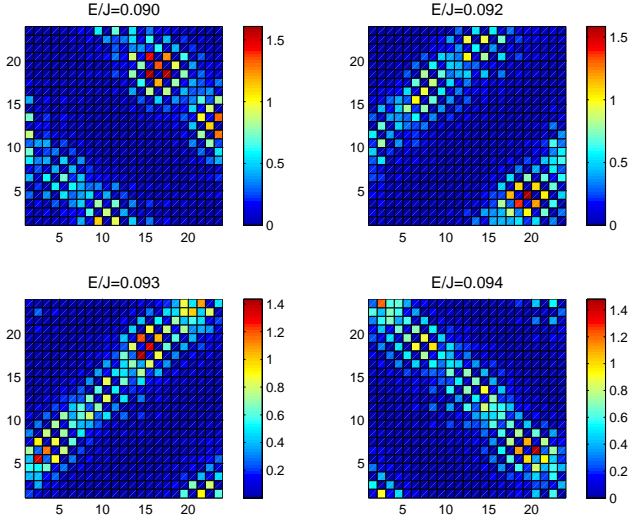} \caption{Low-lying eigenstates, $|v_{ni}|^2$, for
two disorder realizations ((a)-(b) and (c)-(d)) for several
eigenenergies.}
\end{figure}

Additional comparison can be made between the experiment and the
mean-field result. In Ref. \cite{pan}, the authors plot the
integrated local density of states vs. the local gap. Within the
scattering of the data the result follows a monotonic trajectory
implying a larger gap for a smaller $|b_i|^2$ and vice versa.
Figure 5 is such a plot from our mean-field theory. (In making
this plot we included the local density of states for all
24$\times$ 24 sites in two disordered samples.) The result agrees
with the experimental findings
qualitatively. 

\begin{figure}[ht]
\hskip -0.2cm \centering \epsfxsize=5cm \epsfysize=3.7cm
\epsfbox{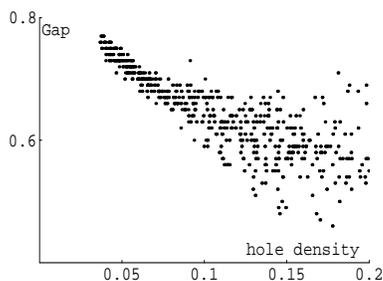}
\caption{Local hole density vs. local peak-to-peak conductance plot.}
\end{figure}

Before closing a caveat is in order. One aspect of our mean-field
result disagrees with the experimental findings -- our conductance
curves for larger gaps show a taller peak while the experimental
finding is  the reverse. It is likely that this discrepancy is due
to the omission of quantum fluctuation of holon phase ($\phi_i$)
in our calculation.

It has recently been reported that after thermal annealing, the
inhomogeneities of the BSCCO surface disappears. Such result
raises question as to whether the observed spectral inhomogeneity
is an intrinsic  bulk property. The point of view we take in this
paper is that even if the surface inhomogeneity is not intrinsic
it still tells us an important information, i.e., the existence of
a situation where the gap varies on nanometer length scale. We
argue that such nanometer-scale variation suggests that the
pairing in cuprates occurs essentially in the spin
sector.\cite{anderson,kivelson}

{\bf Acknowledgment}\rm$~$ We wish to thank Seamus Davis, Eric
Hudson, Kristine Lang, Vidya Madhavan, Joe Orenstein, Shuheng Pan,
and Ziqiang Wang for valuable discussions. We are particularly
grateful to S.H. Pan and S. Davis' group for sharing their data
with us prior to publication. We are indebted to Ziqiang Wang for
suggesting us to look into this problem.  DHL is supported in part
by NSF grant DMR 99-71503. QHW is supported by the National
Natural Science Foundation of China and the Ministry of Science
and Technology of China (NKBSF-G19990646), and in part by the
Berkeley Scholars Program.

\widetext

\begin{references}
\bibitem{note} In the real material, each copper-oxygen plane is
influenced by all the dopant layers, not just one, through
long-range interaction. Therefore $d$ in this paper should be
regarded as an effective distance.
\end{references}
\end{document}